\documentclass[11pt,prd,aps,tightenlines,notitlepage,nofootinbib,preprintnumbers,letterpaper]{revtex4} 
\pdfoutput=1
\usepackage{epsfig}
\usepackage{amsfonts}
\usepackage{amsmath}
\usepackage{graphicx}
\usepackage{color}
\usepackage{amssymb,amsmath,hyperref,slashed,color,graphicx,bm}
\usepackage[T1]{fontenc}

\hypersetup{colorlinks,citecolor= nicegreen,linkcolor= nicered}
\definecolor{nicered}{rgb}{0.7,0.1,0.1}
\definecolor{nicegreen}{rgb}{0.1,0.5,0.1}

\newcommand{\boldgreek}[1]{{\mbox{\boldmath{$#1$}}}}

\newcommand{\beq}{\begin{eqnarray}}
\newcommand{\eeq}{\end{eqnarray}}

\begin{document}

\title{Stable Bound States of Asymmetric Dark Matter}
\author{Mark B. Wise}
\author{Yue Zhang}
\affiliation{ Walter Burke Institute for Theoretical Physics, California Institute of Technology, Pasadena, CA 91125}

\begin{abstract}
The simplest renormalizable effective field theories with asymmetric dark matter bound states contain two additional gauge singlet fields one being the dark matter and the other a mediator particle that the dark matter annihilates into. We examine the physics of one such model with a Dirac fermion as the dark matter and a real scalar mediator. For a range of parameters the Yukawa coupling of the dark matter to the mediator gives rise to stable asymmetric dark matter bound states. We derive properties of the bound states including nuggets formed from $N\gg1$ dark matter particles. We also consider the formation of bound states in the early universe and direct detection of dark matter bound states. Many of our results also hold for symmetric dark matter.
\end{abstract}

\preprint{CALT-TH-2014-145}
\maketitle

\section{Introduction}

An attractive idea for the origin of the cosmological dark matter (DM) density relies on a primordial DM asymmetry 
that prevents the near complete annihilation of the DM particles with their anti-particles resulting in the observed relic density. 
DM that arises from this mechanism is called asymmetric dark matter (ADM)~\cite{Zurek:2013wia}.
If DM is asymmetric then the universe today is composed of DM particles with their antiparticles absent. 
An interesting possibility that can impact the properties of DM relevant for direct and indirect detection 
and influence its distribution in galaxies is the presence of stable bound states of DM particles. 
For ADM these bound states involve DM particles (but not DM anti-particles). 
The cosmology of ADM bound states has been explored previously in various contexts, 
such as the dark atom models with two species of ADM~\cite{Kaplan:2009de, Cline:2012is, CyrRacine:2012fz, Fan:2013tia, Petraki:2014uza},
and strongly interacting non-abelian hidden sector models~\cite{Alves:2009nf, An:2009vq, Detmold:2014qqa, Krnjaic:2014xza}.

In the Standard Model (SM) there is no particle that can be the DM. Minimality can be a useful guiding principle when considering speculative extensions of the SM. The purpose of this paper is to study one of the two minimal renormalizable extensions of the SM with stable ADM bound states. 
We will not discuss the mechanism that generates the primordial DM asymmetry. Rather we focus on the low energy effective theory well below the scale where the primordial asymmetry is generated. In the model we study, the DM is a Dirac fermion $\chi$ with no standard model quantum numbers. To facilitate annihilation of $\chi$ particles with their anti-particles $\bar\chi$ in the early universe, we introduce a Yukawa-coupling of $\chi$ to a real scalar field $\phi$ lighter than the DM. 

Unlike $U(1)$ gauge boson exchange\footnote{DM interacting via dark $U(1)$ gauge boson exchange was considered in for example Refs.~\cite{Ackerman:mha, Feng:2009mn, Zhang:2014wra}.}, scalar exchange is always attractive among particles (or antiparticles), and so for a range of model parameters stable ADM bound states occur\footnote{Having additional scalars aggravates the hierarchy problem of the standard model but in this case since the scalar is related to the DM density there may be environmental reasons for it being light compared with the Planck or GUT scales.} in the model we consider. The new scalar $\phi$  mixes with the Higgs boson, allowing it to decay to SM particles, and mediate interactions between ADM and SM particles that lead to a possible signal in direct detection experiments. 
We discuss the spectrum of non-relativistic ADM bound states from two-body to multi-particle bound states, and the production of bound states in the early universe. 
We find regions of parameter space where most of the DM particles do not reside in bound states, and regions where most of the DM does. The border between these two regions occurs (roughly speaking) when the Yukawa coupling of $\phi$ to the dark matter is large enough and  the $\phi$ mass is comparable to the binding energy of the two body ground state.\footnote{The other minimal renormalizable extension of the standard model with stable ADM bound states has scalar dark matter.} 

The low energy effective theory of ADM we consider must be (approximately) invariant under global $U(1)_{\chi} $ transformations, $\chi \rightarrow {\rm e}^{i\alpha}\chi$. Like the ADM density the baryon density of the universe may arise from a primordial asymmetry, in that case in $B-L$.  For  the baryon asymmetry of the universe, the (approximate) conservation of $B-L$ in the low energy effective theory, i.e., the SM, below the scale where the primordial asymmetry is generated is a consequence of gauge invariance and the particle content, since no renormalizable interactions between SM particles violate $B-L$.\footnote{Violation of this global symmetry occurs at dimension five through the operators that are responsible for neutrino masses.} Naively this is not true for the ADM model we consider, global $U(1)_{\chi}$ invariance is not an automatic consequence of the gauge symmetries and particle content. For example, while a Dirac mass term ${\bar \chi} \chi$ preserves  $\chi$ number a Majorana mass term ${\bar \chi^c} \chi$ does not. However, there could be an unbroken discrete gauge subgroup of  $U(1)_{\chi}$ that forbids the Majorana mass term resulting in a low energy effective theory, where the gauge symmetries are enough to ensure the renormalizable couplings in the low energy theory are invariant under $U(1)_{\chi}$. 

Even in models where the DM is not asymmetric, some of the results of the paper may be applicable. Scalar exchange could give rise to stable bound states of DM particles and stable bound states of anti-DM particles. This will have an impact in cosmology and for direct detection. For a range of parameters, such bound states occur in for example~\cite{MarchRussell:2008tu, Feng:2009hw, Buckley:2009in,Tulin:2013teo}.

\section{The Simplest Fermionic ADM EFT with Bound States}

The most minimal renormalizable low energy effective theory for ADM contains complex scalar DM that annihilates via the Higgs portal in the early universe~\cite{Cheung:2013dca}. However, it is not difficult to show that in this case there is not a robust region of parameter space where non-relativistic ADM bound states exist. Demanding the existence of stable non-relativistic ADM bound states in a renormalizable model requires one more degree of freedom than this minimal model.

We focus on a renormalizable model with Dirac fermionic ADM $\chi$ and a real scalar mediator $\phi$,
\begin{eqnarray}\label{model}
\mathcal{L} &=&i\bar \chi \!\not\!\partial \chi - m_\chi \bar \chi \chi - \bar \chi (g_\chi + i g_5 \gamma_5) \chi \phi + \frac{1}{2} (\partial \phi)^2 - \frac{1}{24}\lambda_{4\phi} \phi^4 - \frac{1}{6}\lambda_{3\phi} \phi^3 - \frac{1}{2} m_\phi^2 \phi^2 \nonumber \\
& -& \mu_{\phi h} \phi( H^\dagger H-v^2/2) - \frac{1}{2}\lambda_{\phi h} \phi^2 (H^\dagger H-v^2/2) - V(H) \ ,
\end{eqnarray}
where $H$ is the SM Higgs doublet, and $V(H)$ is the usual Higgs potential. The vacuum expectation value for the neutral component of the Higgs doublet is $ v/\sqrt{2}$, and $v=246\,$GeV. We have shifted the scalar field $\phi$ so that it has no vacuum expectation value.
After electroweak symmetry breaking, $\phi$ picks up a small mixing with the SM Higgs boson $h$, thus allowing it to decay into SM particles, and mediate the DM interaction to the SM.
This simple model can already have a very complicated spectrum of bound states with important implications for cosmology and direct detection.

With the above interactions, we calculate the cross section for $\chi\bar\chi$ annihilation into the lighter mediator during thermal freeze out. In the region of parameter space with bound states $m_{\phi}$ is much less than $m_{\chi}$. Neglecting $m_{\phi}/m_{\chi}$, $\mu_{\phi h}/m_{\chi}$, $\lambda_{3\phi}/m_\chi$, and expanding to order $v^2$ the annihilation cross section is,
\begin{eqnarray}
\langle\sigma v\rangle_{anni} = \left[\frac{3\pi}{8m_\chi^2} \alpha_\chi^2 v^2 + \frac{2\pi}{m_\chi^2} \alpha_\chi \alpha_I + \frac{\pi}{24m_\chi^2} \alpha_I^2 v^2\right]  \ ,
\end{eqnarray}
where $v$ is the relative velocity between $\chi, \bar \chi$, given by $m_\chi v^2 \simeq 6T$, 
and $\alpha_\chi\equiv g_\chi^2/(4\pi)$, $\alpha_I \equiv g_5^2/(4\pi)$. 
We work to lowest order in perturbation theory neglecting the Sommerfeld enhancement factor which is not very important at freeze out~\cite{MarchRussell:2008tu, ArkaniHamed:2008qn}.
Roughly speaking, asymmetric DM needs a larger annihilation rate than that of the WIMP, $\langle\sigma v(T\simeq m_\chi/26)\rangle_{anni}  > 3 \times 10^{-26} {\rm cm^3/s}$~\cite{Cirelli:2005uq, Graesser:2011wi}.
With this condition, the symmetric component of DM can annihilate efficiently, and today's relic density is dictated by the initial DM asymmetry.

\begin{figure}[t]
\includegraphics[width=1.0\columnwidth]{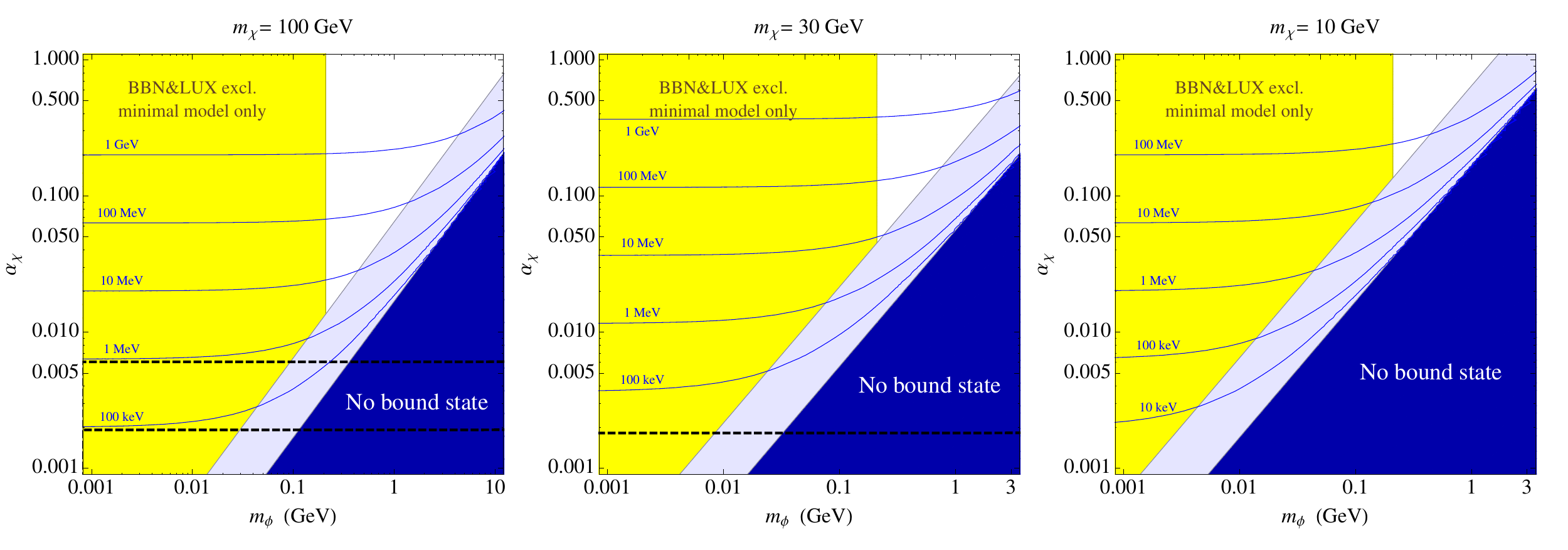}
\caption{Ground state binding energy (in blue curves) of two-body DM bound state in the $\alpha_\chi-m_\phi$ space, for DM mass equal to 1\,TeV (left), 100\,GeV (middle) and 10\,GeV (right), respectively. No two-body bound state exists in the dark blue region, and only one ($1s$) bound state exists in the lighter blue region. The yellow region is not consistent with BBN and LUX constraints in the minimal model described by Eq.~(\ref{model}). The black dashed curves are lower bound on $\alpha_\chi$ from having a large enough annihilation rate for the ADM, in the case $\alpha_I=0$ (upper) and $\alpha_I=\alpha_\chi$, respectively.}\label{bind}
\end{figure}

For $\chi\bar\chi$ annihilation, the contribution from the coupling $g_\chi$ is velocity suppressed but not the contribution from interference between $g_\chi$ and $g_5$. For non-relativistic interactions between DM particles the influence of $g_5$ is suppressed. In  Fig.~\ref{bind} the dashed lines are the limits on $\alpha_{\chi}$ assuming $\alpha_I=0$ (upper dashed line) and $\alpha_\chi=\alpha_I$ (lower dashed lines) that arise from demanding that enough annihilations take place.

In order to be consistent with Big-Bang Nucleosynthesis (BBN), $\phi$ has to decay before a second or so. 
In this model, $\phi$ decays to SM particles via mixing with the Higgs boson $h$, so its decays are similar to those of the Higgs boson~\cite{Gunion:1989we}.
For $m_\phi$ below twice of the pion mass, its decay rate is
\begin{eqnarray}
\Gamma_\phi &=& \left\{ \frac{G_F m_e^2 m_\phi}{4\sqrt{2} \pi} \left( 1- \frac{4m_e^2}{m_\phi^2} \right)^{{3}/{2}} \Theta(m_\phi-2m_e) 
+ \frac{G_F m_\mu^2 m_\phi}{4\sqrt{2} \pi} \left( 1- \frac{4m_\mu^2}{m_\phi^2} \right)^{{3}/{2}} \Theta(m_\phi-2m_\mu) \right. \nonumber \\
& &\left. + \frac{G_F \alpha^2 m_\phi^3}{128\sqrt{2} \pi^3} \left[ A_1 + 2\times 3 \left( \frac{2}{3} \right)^2 A_{{1}/{2}} + 3 \left( \frac{-1}{3} \right)^2 A_{{1}/{2}} \right]^2  \right\} \frac{\mu_{\phi h}^2 v^2}{m_h^4} \ ,
\end{eqnarray}
where the second row is for $\phi\to \gamma\gamma$ decay and we have taken into account of the $W$-loop ($A_1$ term) and the heavy quark loops ($A_{1/2}$ term) from $t,b,c$, 
and $A_1 = -7$, $A_{{1}/{2}} = 4/3$~\cite{Carena:2012xa}. The $\phi\to\mu^+\mu^-$ decay dominates above the two-muon threshold.
For heavier $\phi$, hadronic decay channels open, and the dimuon branching ratio reduces to around 10\%~\cite{Clarke:2013aya}, for $\phi$ mass up to a few GeV.

\begin{figure}[t]
\includegraphics[width=0.6\columnwidth]{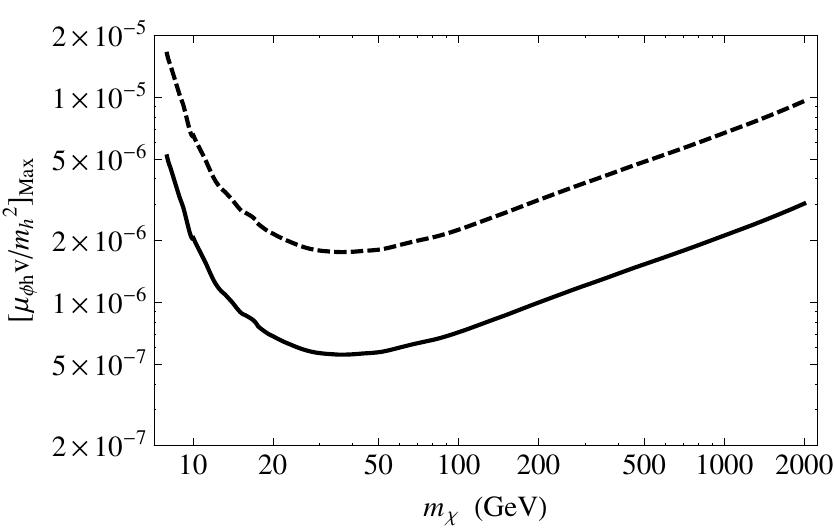}
\caption{Upper bound on the $\phi-h$ mixing parameter $\mu_{\phi h} v/m_h^2$ from dark matter direct detection (LUX), assuming all DM are free particles today. The solid and dashed curves corresponds to $\alpha_\chi=0.1$ and 0.01, respectively. We fix $m_\phi=500\,$MeV.}\label{dd}
\end{figure}

The $\phi-h$ mixing can also mediate the interaction between DM and SM particles, and lead to DM direct detection signals.
The direct detection cross section for $\chi$ scattering on a nucleon is~\cite{Fedderke:2014wda}
\begin{eqnarray}
\label{directdetect}
\sigma_{\rm SI} \simeq \frac{4 \alpha_\chi f^2 m_N^4 \mu_{\phi h}^2}{m_h^4 m_\phi^4} \ ,
\end{eqnarray}
neglecting the $q^2$ in the $\phi$ propagator, where $q$ is the momentum transfer, and $f\simeq 0.35$~\cite{Giedt:2009mr}. 
Fig.~\ref{dd} shows the constraints on the $\phi-h$ mixing from the DM direct detection by the LUX collaboration~\cite{Akerib:2013tjd} for given $\phi$ mass and $\alpha_\chi$.

We find that combining the LUX and BBN constraints generically requires the $\phi\to\mu^+\mu^-$ decay channel to be open, i.e., $m_\phi\gtrsim 210\,$MeV. The regions of parameter space that are excluded by direct detection and consistent with BBN are shaded yellow in Fig.~\ref{bind}. This eliminates a large part of the available parameter space. More complicated models with an enlarged dark sector may evade this constraint. See the appendix for an example.

The scalar field $\phi$ can mediate self interactions between two free DM particles, and is constrained by the bullet cluster observation~\cite{Randall:2007ph} which requires
$\sigma_T/m_\chi < 1.25~ {\rm cm}^2/{\rm gram}$. For some of the  parameter space of interest to this study, {\it i.e.}, $\alpha_\chi m_\chi >m_\phi$,  the Born level cross section $\sigma_T/m_\chi = 4\pi \alpha_\chi^2 m_\chi/m_\phi^4$ is not valid  and quantum mechanical effect become relevant. Unless resonantly enhanced, the quantum regime the cross section is typically smaller than the naive Born estimate~\cite{Tulin:2013teo}. With this knowledge, we find that typically the bullet cluster constraint is only important in the region of parameter space with a light mediator $m_\phi<0.5\,$GeV and strong coupling $\alpha_\chi>0.3$.

\subsection{Two-body Bound States}

The exchange of a scalar field $\phi$ gives an attractive force among the ADM particles. For sufficiently light $\phi$ bound states occur. Because the $\phi$ force is always attractive, multi-particle ADM bound states ($N\geq2$) are also expected.

In this section, we discuss two-body bound states. The states containing more than two DM will be discussed in the following subsection.

For two-DM-particle system, the non-relativistic Hamiltonian describing two DM interaction is
\begin{eqnarray}\label{HNR}
H &=& -\frac{{\boldgreek \nabla}_{\rm cm}^2}{4m_\chi} -\frac{{\boldgreek \nabla}^2}{m_\chi} - \frac{\alpha_\chi}{r} e^{-m_\phi r} + H_{int} \ , 
\end{eqnarray}
where ${\bf x}_{\rm cm}$ is the center of mass coordinate, and ${\bf r}$ is the relative position of two $\chi$'s.
$H_{int}$ is the interaction for on-shell $\phi$ creation/annihilation, which controls transition rates, that involve $\phi$ particle emission and absorption. The other terms in Eq.~(\ref{HNR}) control the spectra and wave functions of the two-body bound states and scattering states in the non-relativistic limit.

The non-relativistic two-body bound state problem with a Yukawa potential has been solved numerically in~\cite{yukawa} (see also~\cite{Shepherd:2009sa}).
In order for a bound state to exist, the screening length ($\sim$\,$1/m_\phi$) must be large enough compared to the size of the corresponding wave function ($\sim$\,the Bohr radius).
The condition for having at least one bound state (the $1s$ state) is
\begin{eqnarray}\label{boundcondition}
\frac{\alpha_\chi m_\chi}{2 m_\phi} > 0.8399 \ .
\end{eqnarray}

In Fig.~\ref{bind}, we plotted contours of fixed $1s$ state binding energy as thin blue curves in the $\alpha_\chi$ versus $m_\phi$ plane, for fixed DM mass in each panel.
The dark blue region does not satisfy the  bound state condition in Eq.~(\ref{boundcondition}). 
In the light blue region, the 1s state is the only bound state. 
Outside these regions, for $m_\phi\ll\alpha_\chi m_\chi$, the $\phi$ force is very close to Coulombic, and the solution approaches a hydrogen-like state, with a Bohr radius
\begin{eqnarray}
a_0 = \frac{2}{\alpha_\chi m_\chi} \ .
\end{eqnarray}
In this case, the ground state wave function and binding energy are, approximately,
\begin{eqnarray}\label{hydrogenground}
\psi_0(r) = \frac{1}{\sqrt\pi} a_0^{- {3}/{2}} e^{- {r}/{a_0}}, \hspace{1cm} { BE}_0 = \frac{\alpha_\chi^2 m_\chi}{4} \ .
\end{eqnarray}

A two-particle bound state $B_i$ can decay to one with greater binding energy
$B_f$ by real or virtual $\phi$ emission. Real $\phi$ emission is kinematically allowed when the the difference in binding energies $\Delta{ BE}$ between the final and initial states is greater than the mass of the $\phi$. Also bound states can be formed by scattering two $\chi$'s and emitting either a real or virtual $\phi$ and dissociated by scattering the bound state with a $\phi$. The couplings of the $\phi$ to the SM particles are restricted to be very small by current direct detection bounds so we concentrate on real $\phi$ emission and absorption in these processes. Also for simplicity we only consider the $s$-wave bound states. They have the spins of the $\chi$'s combined into a total spin zero state.

In the non relativistic limit the interaction Hamiltonian that is needed for these calculations is
\begin{eqnarray}
\label{nrham}
H_{int} &=& g_\chi \left[\phi\left({\bf x}_{\rm cm} + \frac{{\bf r}}{2}\right) + \phi\left({\bf x}_{\rm cm}-\frac{{\bf r}}{2}\right)\right] 
 \ ,
\end{eqnarray}
where we neglected the $g_5$ term which plays a subdominant role here.
Our calculations are valid as long as the $\chi$'s are non-relativistic but the $\phi$ can be relativistic. 
Excited bound states can decay to lower bound states by $\phi$ emission,
\begin{equation}
\Gamma(B_i \rightarrow B_f +\phi)=8 \alpha_{\chi} k |G_{if}(k)|^2 \ ,
\end{equation}
where $k=\sqrt{\Delta{ BE}^2-m_{\phi}^2}$ is the magnitude of the $\phi$ momentum and the transition form factor is
\begin{equation}
G_{if}({ k})=\int d^3r e^{-i{\bf k}\cdot{\bf r}/2}\psi_f^*({\bf r})\psi_i({\bf r}) \ .
\end{equation}

We will need, for the cosmology discussion in the forthcoming section, the cross sections for the formation and dissociation of the two-body states via real $\phi$ emission and absorption (see Fig.~\ref{feynmandiagram}).   The relativistic correction to the Hamiltonian  in eq.~(\ref{nrham})
\begin{equation}
\label{relham}
\Delta H_{int}=g_r \left[\phi\left({\bf x}_{\rm cm} + \frac{{\bf r}}{2}\right) + \phi\left({\bf x}_{\rm cm}-\frac{{\bf r}}{2}\right)\right] \left(  \frac{{\boldgreek \nabla}^2}{2m_\chi^2} + \frac{{\boldgreek \nabla}_{\rm cm}^2}{4m_\chi^2} \right)
\end{equation}
is important for  bound state production and dissociation.

We first consider the formation process  in the center of mass frame, $\chi ({\bf p})+\chi (-{\bf p}) \rightarrow B_i+ \phi $. Including the relativistic correction from eq.~({\ref{relham}) the cross section is expressed in terms of the form factor.
\begin{eqnarray}\label{form}
F_i({\bf p}, {\bf k}) =\int d^3{\bf r} \psi_i^*({\bf r}) (e^{i{\bf k}\cdot {\bf r}/2}+e^{-i{\bf k}\cdot {\bf r}/2}) \left( 1 + \frac{{\boldgreek \nabla}^2}{2m_\chi^2} \right) \psi_c ({\bf p}, {\bf r})  \ ,
\end{eqnarray}
where $\psi_i$ is the spatial wave function for the $i$-th s-wave bound state, $\bf k$ is the three momentum of the outgoing $\phi$, and (neglecting $m_{\phi}$)
$\psi_c$ is the Coulomb wave function with fixed incoming momentum ${\bf p}$,
\begin{eqnarray}
\psi_c ({\bf p}, {\bf r}) = e^{\pi/(2 a_0 p)} \Gamma \left(1-\frac{i}{a_0 p} \right) F\left( \frac{i}{a_0 p}, 1, i (p r - {\bf p}\cdot {\bf r}) \right) e^{i {\bf p}\cdot {\bf r}} \ ,
\end{eqnarray}
and $F$ is the confluent hyper-geometric function.

Because the above $\psi_i$ and $\psi_c$ are eigenstates of the same Hamiltonian with different energy eigenvalues, they are orthogonal each other.
Therefore, one has to go to order $p^2/m_\chi^2$ by including the relativistic correction, or to order $k^2/(p^2 + 1/a_0^2)$ in the small $k$ expansion.
Using the identity~\cite{Pearce:2013ola},
\begin{eqnarray}
\int \frac{d^3 {\bf r}}{r} e^{i ({\bf p}-{\bf k)}\cdot {\bf r} - \eta r} F\left( i\xi, 1, i (p r - {\bf p}\cdot {\bf r}) \right) = 4\pi \frac{[k^2 + (\eta -ip)^2]^{-i\xi}}{[({\bf p}-{\bf k})^2 + \eta^2]^{1-i\xi}} \ ,
\end{eqnarray}
we find the squared form factor after averaging over the angle between ${\bf p}$ and ${\bf k}$ to be
\begin{eqnarray}
|F_{\rm eff} (p, k)|^2 &=& \frac{64\pi}{a_0^3} e^{\frac{\pi}{a_0 p}} \left| \Gamma \left(1-\frac{i}{a_0 p} \right) \frac{(\eta - i p)^{-2i\xi}}{(\eta^2 + p^2)^{1-i\xi}} \right|^2  \nonumber \\
&&\hspace{-2.2cm}\times\left[ \frac{\alpha_\chi^2}{4m_\chi^2} - \frac{\alpha_\chi}{3m_\chi a_0} \frac{k^2 p^2}{(p^2 + 1/a_0^2)^3} \left( 1 + \frac{1}{p^2 a_0^2} \right) + \frac{1}{15} \frac{k^4 p^4}{(p^2 + 1/a_0^2)^6} \frac{1}{a_0^2} \left( 1 + \frac{1}{p^2 a_0^2} \right) \left( 23 + \frac{7}{p^2 a_0^2} \right)\right]. \nonumber \\
\end{eqnarray}
Here we have neglected $m_{\phi}$. It is interesting to examine $|F_{\rm eff} (p, k)|^2$ in the limit  of small $p$, {\it i.e.}   $p\ll1/a_0$. Then, using the Sterling's approximation
\begin{equation}
|F_{\rm eff} (p, k)|^2 \rightarrow \left( {128 \pi^2 \over e^4 p} \right) \left[ {\alpha_{\chi}^2 \over 4 m_{\chi}^2}-{k^2 a_0^3 \alpha_{\chi} \over 3 m_{\chi}}+{7 \over 15} k^4a_0^6 \right] \ .
\end{equation}

In the center of mass frame, 
\begin{equation}\label{XSform}
\sigma(\chi ({\bf p})+\chi (-{\bf p}) \rightarrow B_i +\phi)v= 2\alpha_{\chi} k |F_{\rm eff} (p, k)|^2 \ ,
\end{equation}
where energy conservation fixes the magnitude of the final state $\phi$ momentum to be,
\begin{equation}\label{15}
k=\sqrt{ (BE_i+p^2/m_{\chi})^2-m_{\phi}^2} \ .
\end{equation}
It is evident from the above formula that as long as the $\chi$'s are non-relativistic, keeping only the leading $k$ dependence in the form factor is reasonable.

Similarly the cross section for the dissociation process in the lab frame, $ B_i({\bf 0})+\phi({\bf k})\rightarrow \chi+\chi$   is,
\begin{equation}\label{XSdiss}
\sigma(B_i({\bf 0})+\phi({\bf k})\rightarrow \chi+\chi)v={\alpha_{\chi} \over E_k}m_{\chi}p  |F_{\rm eff} (p, k)|^2 \ ,
\end{equation}
where $p$ is the magnitude of the relative momentum of the final state $\chi$'s, {\it i.e.} ${\bf p}=({\bf p}_1-{\bf p}_2)/2$  with ${\bf p}_1$ and ${\bf p}_2$ are the three momenta of the two final state $\chi$'s. $E_k=\sqrt{k^2+m_{\phi}^2}$ is the energy of the incoming $\phi$. Energy conservation determines the magnitude of the relative $\chi$ momentum to be,
\begin{equation}
p=\sqrt{m_{\chi}(E_k-BE_i)} \ ,
\end{equation}
where $BE_i$ is the binding energy of the two body bound state $B_i$.

\begin{figure}[t]
\includegraphics[width=0.9\columnwidth]{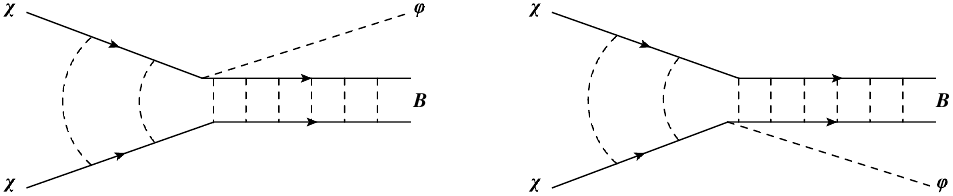}
\caption{Feynman diagrams for two-body ADM bound state formation (dissociation) with an on-shell $\phi$ emission (absorption).}\label{feynmandiagram}
\end{figure}

If some of the DM is in the two-body bound ground state today the direct detection predictions may be modified. The momentum transfer associated with direct detection is less than $1/a_0$ so under most circumstances the scattering is coherent, the form factor suppression negligible and the cross section is just 4 times the single $\chi$ cross section in Eq.~(\ref{directdetect}). An exception occurs when the two body scattering length is very large. At zero momentum the (spin zero) $\chi-\chi$ scattering cross section $\sigma(\chi \chi \rightarrow \chi \chi)=4 \pi a^2$, where $a$ is the spin zero scattering length. If there is a two body bound state (or resonance) very near threashold ({\it i.e.} zero binding energy) then the scattering length $a$ is very large\footnote{This is the origin of the enhancements in \cite{MarchRussell:2008tu}, \cite{Buckley:2009in} and \cite{Tulin:2013teo}.}. In this case there are universal  properties at low momentum for few body $\chi$ systems. This is familiar from the effective range expansion for two body nuclear physics processes ({\it e.g.} $n+p \rightarrow d +\gamma$) and was applied to DM direct detection in \cite{Laha:2013gva}.

\subsection{Dark Matter Nuggets}

As mentioned above, the $\phi$ force among DM particles is always attractive. 
Therefore, for small enough $\phi$ mass we expect DM to have $N>2$ particle bound states (nuggets).
For sufficiently large $N$, we assume the DM nugget can be described as non-relativistic degenerate Fermi gas. 
Undoubtably this is a gross simplification of the dynamics. One  expects more complicated phenomena
like pair formation~\cite{Shankar:1993pf} to occur and impact the equation of state for the DM. However, our main purpose here is not a quantitative analysis of the spectrum and properties of multi-particle DM bound states but rather to argue that such states exist and that they are probably small enough that in direct detection experiments the scattering is coherent and gives rise to a cross section that grows as $N^2$.
This will also affect the capture rate for DM by neutron stars~\cite{Kouvaris:2011gb, McDermott:2011jp, Bell:2013xk, Bramante:2013nma}. The capture rate may be enhanced by the self interactions of ADM~\cite{Kouvaris:2011gb, Guver:2012ba}.

We first discuss this multi-particle bound state problem using a heuristic approach where the DM density is constant and then a more quantitative approach that relies on hydrostatic equilibrium. In expressions for physical quantities the scaling with $N$, $m_{\chi}$ and $\alpha_{\chi}$ is the same in the heuristic and hydrostatic equilibrium approaches.

We restrict our attention to the non-relativistic regime so we can neglect relativistic corrections to the potential energy, for example from the Darwin term. Also we will find that the size of the state decreases with $N$ so that we can neglect $m_{\phi}$ replacing the Yukawa potential by a Coulomb potential.

\subsubsection{Heuristic Approach}

Assuming a constant density of $\chi$ particles filling the momentum levels up to the Fermi momentum $p_F$ the number density of $\chi$ particles is
\begin{equation}
n= {p_F^3 \over 3 \pi^2} \ .
\end{equation}
For a spherical volume of radius $R$, the total number of particles is $N=4 p_F^3R^3 /(9 \pi)$.
The kinetic energy $KE$, and potential energy $PE$, of the $\chi$ particles expressed as a function or the total number of particles $N$ and the radius $R$ are
\begin{equation}
KE= {2 \pi^{2/3} \over 15 R^2m_{\chi}} N^{5/3}\left(9 \over 4\right)^{5/3}, \hspace{1cm}
PE=-{ 3N^2 \alpha_{\chi} \over 5 R} \ .
\end{equation}
Minimizing the total energy $E=KE+PE$ with respect to $R$ at fixed $N$ determines the radius to be,
\begin{equation}
\label{nuggetsize}
R=\left({9 \pi \over 4}\right)^{2/3}{ 1 \over N^{1/3}\alpha_{\chi}m_{\chi}} \ .
\end{equation}
Notice that the volume of the nugget decreases as $1/N$ and if $m_{\phi}a_0\ll1$, then we also have
for large $N$ that $m_{\phi}R\ll1$. Hence as long as two body bound states exist  the Yukawa potential can be treated as Coulombic for nuggets. 

At the value of $R$ in Eq.~(\ref{nuggetsize}) the kinetic  and potential energies become,
\begin{eqnarray}
KE
\simeq 
0.08\left(\alpha_{\chi}^2 m_{\chi} N^{7/3}\right) \ , \hspace{1cm}
PE
\simeq 
-0.16 \left(\alpha_{\chi}^2 m_{\chi} N^{7/3}\right) \ .
\end{eqnarray}
For large $N$ the nuggets have a binding energy that is of order $N^{4/3}$ times the binding energy of $N/2$ two body bound states $BE_0$.

There are a number of conditions that must be satisfied to apply even the crude approximations we have made. Firstly as $N$ increases the Fermi momentum increases and the system eventually becomes relativistic. Demanding that $p_F/m_{\chi}\ll 1$ implies that
\begin{equation}
N\ll \left({9\pi \over 4}\right)^{1/2} \alpha_{\chi}^{-3/2} \simeq 2.7 \left({1 \over \alpha_{\chi}}\right)^{3/2} \ .
\end{equation}
Determining the properties of the nugget using classical methods is valid for
$p_F R\gg1$. This implies that
\begin{equation}
N\gg\left( {9 \pi \over 4} \right)^{-1/3}  \simeq  0.52 \ .
\end{equation}

In the presence of a background number density for $\chi$'s, the Yukawa coupling of the scalar induces a density dependent tadpole and for large enough $R$ a scalar expectation value. However, neglecting $m_{\phi}$ and $\lambda_{3\phi}$ and treating $\lambda_{4\phi}$ as order unity, we find that these effects are subdominant compared to those we have included, provided $N\gg\sqrt{1/\alpha_{\chi}}$.

\subsubsection{Hydrostatic Equilibrium}

For a non-relativistic degenerate Fermi gas the equation of state  relating the pressure density $p$ to the number density $n$  is
\begin{eqnarray}\label{eos}
p = Kn^{5/3} \ ,
\end{eqnarray}
where $K = 5^{-1} {3^{2/3} \pi^{4/3}}{ m_\chi^{-1}}$,  and $n$ is the number density of $\chi$ particles.
For a stable solution, the Fermi pressure is balanced by the attractive $\phi$ force among the particles. 
When ${1}/{m_{\phi}}$ is larger than the size of the nugget, the attractive force is Coulomb-like. 
In this case, the hydrostatic equilibrium equation is
\begin{eqnarray}
\frac{1}{r^2} \frac{d}{dr} \left( \frac{r^2}{n} \frac{dp}{dr} \right) = -{4\pi \alpha_\chi n} \ .
\end{eqnarray}
Together with the above equation of state, Eq.~(\ref{eos}), this equation can be solved for $N$ DM particles and it has a finite-size solution, with
\begin{eqnarray}
R = \frac{(3\pi)^{2/3} 2^{-7/3}}{N^{1/3} \alpha_\chi m_\chi} (\xi_1^2 |\theta'(\xi_1)|)^{1/3} \cdot \xi_1 \simeq \frac{4.5}{N^{1/3} \alpha_\chi m_\chi} \ ,
\end{eqnarray}
where $\theta(\xi)$ is the solution to the Lane-Emden equation with index $n =3/2$, and $\xi_1=3.65$, $\xi_1^2 |\theta'(\xi_1)|=2.71$~\cite{book} . 
The Fermi momentum near the center of the nugget is
\begin{eqnarray}
(p_F)_c = \frac{2^{2/3} N^{2/3} \alpha_\chi m_\chi}{3^{1/3} \pi^{4/3} (\xi_1^2 |\theta'(\xi_1)|)^{2/3}} \simeq 0.1 N^{2/3} \alpha_\chi m_\chi \ .
\end{eqnarray}
For this description to apply, there are consistency conditions.
\begin{itemize}
\item Non-relativistic condition: $(p_F)_c \ll m_\chi$ requires $N \ll \left({0.1 \alpha_\chi} \right)^{-3/2}$. For $\alpha_\chi \leq0.1$, the right hand side is of order $10^3$ or larger.
\item Classical description: $(p_F)_c R \gg 1$ requires $N \gg 6.4$\ .
\item Long range force condition: $m_\phi R \ll 1$ requires $m_\phi \ll {N^{1/3}} \alpha_\chi m_\chi/{4.5}$\ .
\end{itemize}
Large $N$ nuggets are smaller than the the two-body bound ground state, while the momentum of the DM inside the nugget is much larger than in the two-body case. Therefore, as long as the two-body bound state exists, the screening effect due to $m_\phi$ can be neglected.
Note that nuggets with large $N$ can exist even for $m_\phi a_0>1$ where the two-body bound states do not occur.

\medskip
Within the degenerate Fermi gas picture, for very large $N$, a non-relativistic description is no longer valid.  Other interactions we have not included become important in the analysis of such systems. However, it seems plausible that relativistic bound states exist.

\medskip
As we noted before, since the size of DM nuggets shrinks with $N$ we expect their direct detection scattering cross section to be coherent and be proportional to $N^2$.

\section{Cosmology}

In this section, we study the formation of bound states in the early universe. In general, there are two stages in the evolution of the universe when bound states may form most efficiently: 1) shortly after the DM freeze out when the ADM number density is still high, 2) at a later stage where structure growth has gone non-linear and the DM density can be locally large. Here we focus on 1).

We perform a calculation of two-body bound state production in stage 1), taking into account
two competing processes, formation $\chi\chi\to B \phi$, and dissociation $B \phi\to \chi\chi$.
The fraction of DM in bound states depends on the interplay between these two rates and the Hubble parameter.
Our goal here is to understand two-body bound state production.
To calculate the formation of bound states with more than two particles, we need to know the binding energies and wave functions of those states.

For convenience, in this section we fix $\mu_{\phi h} v/m_h^2=10^{-7}$ and take $m_\phi>2m_\mu$. 
These values are consistent with constraints from DM direct detection for $\alpha_\chi<1$.
In this region of parameter space, $\phi$ decay is dominated by hadronic and the two muon-final state.
Neglecting threshold effects, the two-muon contribution gives the bound,
\begin{eqnarray}
\tau_\phi < 10^{-2} \,{\rm sec} \left( \frac{1\,\rm GeV}{m_\phi} \right) \ .
\end{eqnarray}
Recall that then the universe is $10^{-2}$ second old, its temperature is $\sim 10\,$MeV.

When $t<1/\tau_\phi$ in the early universe there is a plasma of $\phi$ particles which couples to the DM.
In order to calculate the averaged bound state formation and dissociation rates in this plasma, 
we need to know the energy/momentum distributions of $\chi$ and $\phi$.
At very high temperature, the mediator $\phi$ was in thermal equilibrium with SM fermions via the Higgs boson exchange, $\phi \phi\leftrightarrow f\bar f$. For $\lambda_{\phi h}\sim \mathcal{O}(1)$ and $m_\phi\lesssim1\,$GeV, such interactions freeze out at temperature equal to 1 GeV or so, slightly below the charm quark threshold.
Afterwards, $\phi$ can only remain in chemical equilibrium with itself through the $2\leftrightarrow3$ scattering $\phi\phi \leftrightarrow \phi\phi\phi$, with the $\lambda_{3\phi}, \lambda_{4\phi}$ couplings. 
This allows it to have its own temperature $T_\phi$, which satisfies
\begin{eqnarray}
\frac{T_\phi}{T_\gamma} \simeq
\left\{ \begin{array}{cr}
1,& \hspace{0.5cm} T_\gamma>1\,{\rm GeV} \\
\left[ {g_*(T_\gamma)}/{g_*({\rm 1\,GeV})} \right]^{1/3}, & \hspace{0.5cm} T_\gamma<1\,{\rm GeV}
\end{array}\right.
\end{eqnarray}
where $T_\gamma$ is the photon temperature and $g_*(T_\gamma)$ is the number of relativistic degrees of freedom in the SM at $T_\gamma$.
When the temperature falls below $m_\phi$, the $\phi$ number density becomes Boltzmann suppressed and the $2\leftrightarrow3$ scattering process freezes out, at temperature around an order of magnitude below $m_\phi$.
Another important way to deplete the $\phi$'s is decay. 
For simplicity, we take 
the phase-space distribution of $\phi$ to be
\begin{eqnarray}
f_\phi(E) \simeq e^{- E/T_\phi} e^{-1/(2H\tau_\phi)} \ .
\end{eqnarray}

Second, after the anti-DM $\bar\chi$ are efficiently depleted ($T\lesssim m_\chi/30$), the remaining ADM component $\chi$ can stay in kinetic equilibrium with $\phi$, via the elastic scattering $\chi\phi\to\chi\phi$. For the range of parameters we study, this rate is always larger than the Hubble rate, until $\phi$'s decay away. In this case, the phase-space distribution of DM $\chi$ is
\begin{eqnarray}
f_\chi(p) = C e^{- p^2/(2m_\chi T_\phi)}, \hspace{0.5cm} C=2^{5/2} \pi^{-1/2} \zeta(3) \frac{T_\gamma^3}{(m_\chi T_\phi)^{3/2}} \left( \eta \frac{\Omega_{\rm DM}}{\Omega_b}  \frac{m_p}{m_\chi} \right) \ ,
\end{eqnarray}
where $\eta\simeq6\times10^{-10}$ is the ratio of baryon to photon number in the universe, and $m_p$ is the mass of the proton.

We will use the $\phi$ and $\chi$ distributions described above to calculate the thermal averaged rates.
For the formation rate $\Gamma_{form}$ per particle, we do a thermal integral of the cross section Eq.~(\ref{XSform}) over the incoming $\chi$ momentum.
The thermally averaged dissociation rate $\Gamma_{diss}$ per particle is obtained by integrating Eq.~(\ref{XSdiss}) over the incoming $\phi$ energy.

When both $\Gamma_{form}$ and $\Gamma_{diss}$ are larger than the Hubble rate and only two-body bound states exist,
the dark ionization fraction, $X_d \equiv n_\chi/(2\sum_in_{B_i}+n_\chi)$, (the index $i$ goes over all possible two-body bound states)
satisfies the dark sector counterpart of the Saha equation\footnote{A similar equation could be derived in the nugget case.}
\begin{equation}
\frac{1-X_d}{X_d^2} = \frac{8 \zeta(3)}{\sqrt\pi} \eta \left( \frac{5.4\,\rm GeV}{m_\chi} \right) \left( \frac{T_\phi}{m_\chi} \right)^{3/2} \sum_i e^{BE_i/T} \ .
\end{equation}
The large number of $\phi$'s in the plasma implies that $\Gamma_{diss}\gg\Gamma_{form}$ until the time of $\phi$ decay.
In the end, how many two-body bound states are formed is determined by the comparison of formation rate and Hubble rate at that time.

For simplicity below, we only discuss the formation and dissociation of the ground state. However, 
kinematically, excited states are harder to form and easier to destroy. Therefore, we expect most of the two-body bound state formation to occur in the ground state.
The parameter space can be divided into two regimes.

\begin{itemize}

\item Case A: $m_\phi \gg BE_0$.

In this case, the mediator mass is much larger than the binding energy, 
and the formation process $\chi\chi\to B \phi$ cannot happen unless the two DM are energetic enough.
In other words, for DM in kinetic equilibrium, the temperature of the universe must be large enough.
When the temperature falls below a threshold $T_{th} = (m_\phi-BE_0)/3 \simeq m_\phi/3$, the formation rate becomes exponentially suppressed.
In contrast, with a plasma of $\phi$ the dissociation rate does not shut off until $\phi$ eventually decays, which happens at a temperature lower than $T_{th}$.

This feature is shown as the left panel of Fig.~\ref{BE}, where when $\Gamma_{form}$ falls below $H$, $\Gamma_{diss}$ is still much larger than the Hubble rate $H$. In this case, any bound states that were formed will eventually be ionized back to unbounded DM particles. 
For the same reason, bound state formation when the structure growth becomes non-linear is also suppressed because the DM is more non-relativistic.

\begin{figure}[t]
\includegraphics[width=1.0\columnwidth]{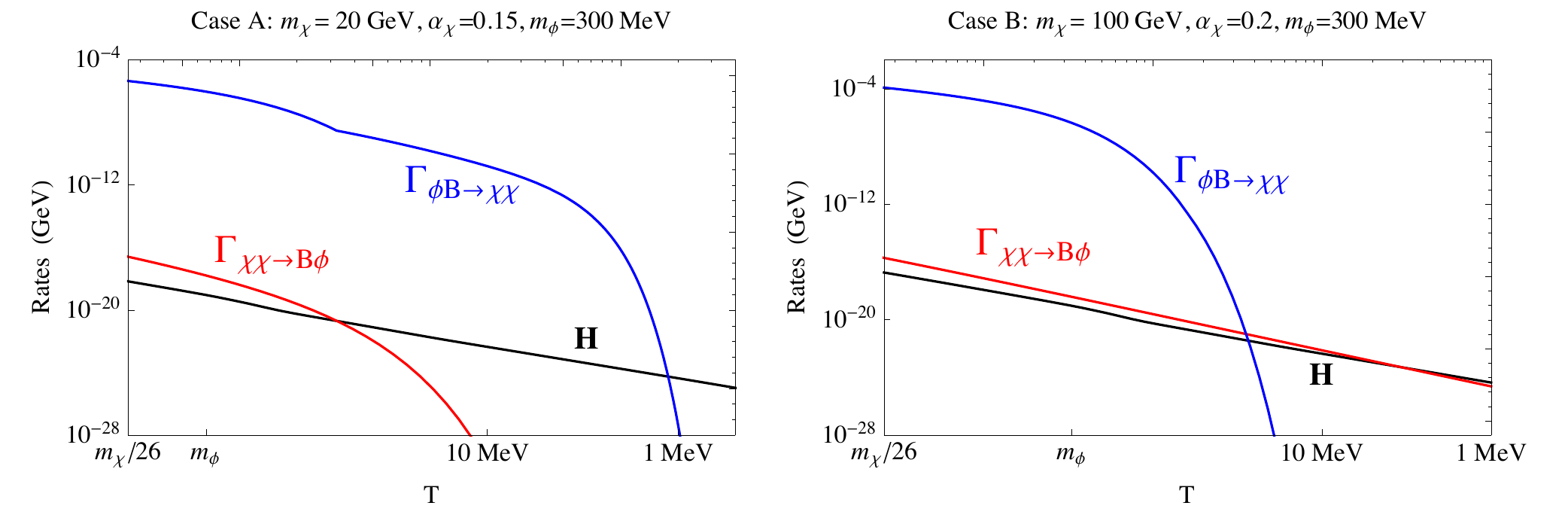}
\caption{Temperature dependence of the two-body ADM bound state formation (red), dissociation (blue) rates in together with the Hubble expansion rate (black).
In both cases, we set $\mu_{\phi h} v/m_h^2 =10^{-7}$ such that both DM direct detection and BBN bounds can be satisfied.}\label{BE}
\end{figure}

\begin{figure}[t]
\includegraphics[width=0.6\columnwidth]{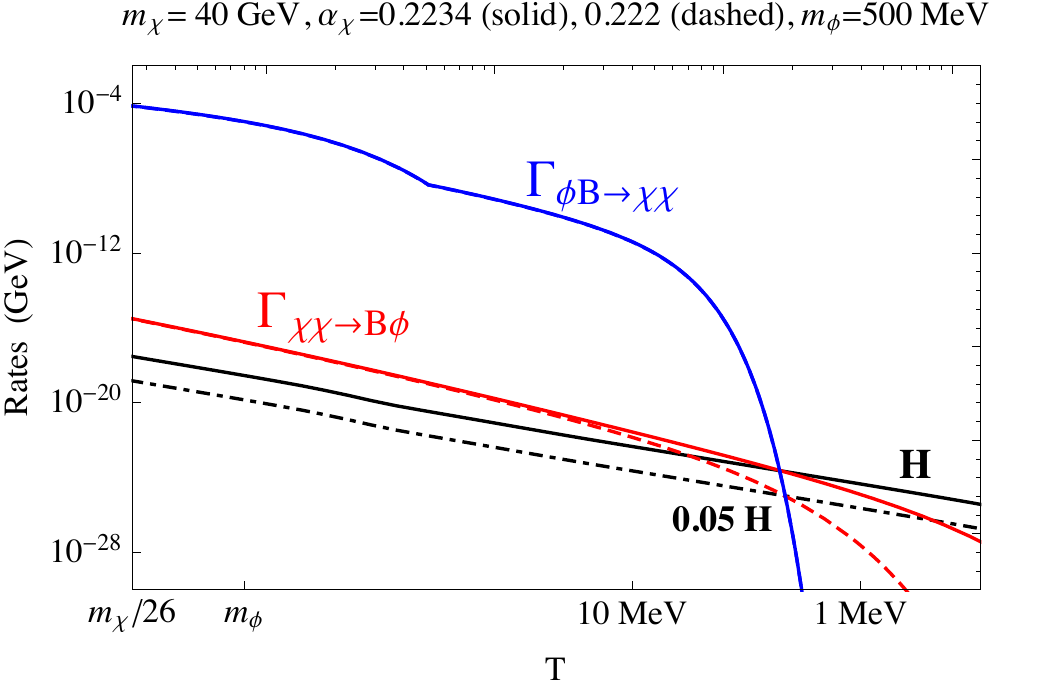}
\caption{Same as Fig.~\ref{BE}, but the parameters are chosen such that at the temperature when $\Gamma_{diss}= H$, the formation rate satisfies $P=\Gamma_{form}/H=1$ (solid), or $P=\Gamma_{form}/H=5\%$ (dashed). }\label{critical}
\end{figure}

\item Case B: $m_\phi \ll BE_0$.

In this case, the binding energy release itself is sufficient to produce an on-shell $\phi$. There is no temperature threshold for bound state formation.
The thermally averaged formation cross section Eq.~(\ref{XSform}) can be simplified in the following two regimes
\begin{equation}
\langle\sigma(\chi+\chi \rightarrow B_0 +\phi)v\rangle \simeq 
\left\{\begin{array}{cr}
{0.5\pi \alpha_\chi^6}/({m_\chi T_\phi}), & \hspace{1cm}T_\phi\gg BE_0 \vspace{0.05in}\\
10 \pi^2 \alpha_\chi^5/\sqrt{m_\chi^{3} T_\phi}, & \hspace{1cm}T_\phi\ll BE_0
\end{array}\right.
\end{equation}
Because $n_\chi\sim T_\phi^3$, the formation rate keeps decreasing as $T_\phi$ drops.
Hence, as a necessary condition for bound state production, there is a lower bound on the coupling constant $\alpha_\chi$. For the formation rate to ever be larger than the Hubble rate it must be larger at $T\sim m_\chi/30$, which implies that,
\begin{equation}\label{smallestalpha}
\alpha_\chi \gtrsim 0.1 \left( m_\chi \over 100\,{\rm GeV} \right)^{1/3} \ .
\end{equation}
If this condition is satisfied, the formation rate can remain greater than the Hubble rate for a long time.
In contrast, there will be a threshold for dissociation. 
At $m_\phi <T< E_B$, there is a suppression in the number of $\phi$ in the plasma that are energetic enough to ionize the bound state. 
Moreover, there is a sharper suppression when $\phi$ begins to decay.

This allows us to have a picture where the ground state formation process is still active ($\Gamma_{form} > H$) when dissociation is suppressed ($\Gamma_{diss}<H$), as shown in the right panel of Fig.~\ref{BE}. In this regime, the two-body ground state can efficiently form.
In this regime, the formation of more than two-body bound states is also expected to be efficient.

\begin{figure}[t]
\includegraphics[width=1.0\columnwidth]{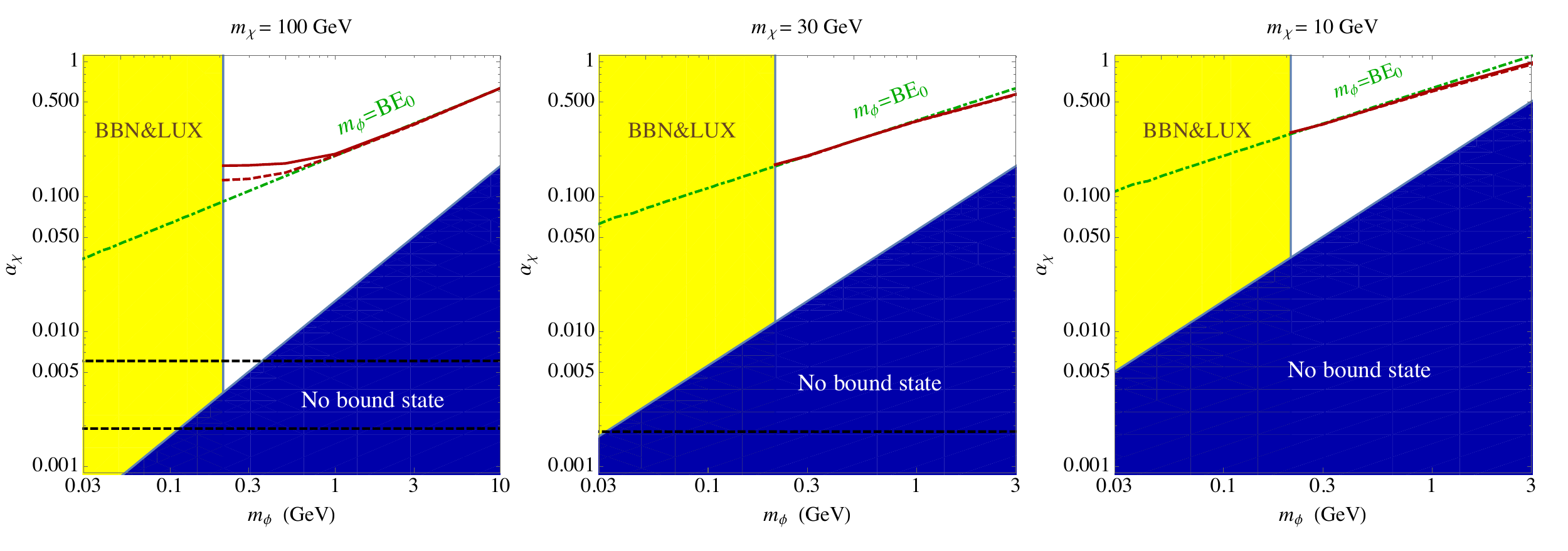}
\caption{Same parameter space as Fig.~\ref{bind}. The thick red curves represent where the critical condition like Fig.~\ref{critical} is reached, 
{\it i.e.}, at the temperature when $\Gamma_{diss}= H$, the formation rate satisfies $P=\Gamma_{form}/H=1$ (solid), or $P=\Gamma_{form}/H=5\%$ (dashed).
The dot-dashed lines are where $m_\phi=BE_0$ is satisfied.}\label{boarder}
\end{figure}

\item The critical case.

In the region that interpolates between the two above limiting cases, for given $m_\phi$, we find it is always possible to arrange the parameters such that 
both formation and dissociation freeze out at similar temperature $T_c$, where 
$\Gamma_{diss}(T_c) = H (T_c)$, but $\Gamma_{form} (T_c)$ is somewhat below $H (T_c)$, as shown in Fig.~\ref{critical}.
Below $T_c$, the dissociation rate is more suppressed because of $\phi$ decay.
In this case, the fraction of DM particles that finally end up in the two-body ground state is roughly,
\begin{eqnarray}\label{Prediction}
P \simeq \frac{\Gamma_{form}}{H}(T_c) \ .
\end{eqnarray}
Since $P$ is smaller than one, we expect the formation of more than two-body bound states to be further suppressed.

\end{itemize}

We summarize the results in the $\alpha_\chi$ versus $m_\phi$ parameter space in Fig.~\ref{boarder}.
The thick red curves are where the above critical condition is satisfied.
Below the thick red curves, almost all the DM ends up as unbound $\chi$ particles.
Above solid (dashed) red curves, most (5\%) of the DM resides in bound states.
These curves lie at $m_\phi$ somewhat above the binding energy $BE_0$ because of the kinetic energy of the $\chi$ particles (see Eq.~(\ref{15})).
Note the thick red curves bend up towards/crossing the $m_\phi=BE_0$ line near the two muon threshold for $\phi$ decay.
This occurs because the $\phi$ decay rate is suppressed which postpones the suppression of dissociation rate, and a larger value of $\alpha_\chi$ is needed to reach the critical point.
In this regime, if the DM is sufficient heavy, the binding energy $BE_0$ can already exceed $m_\phi$ (see the $m_\chi=100$\,GeV case for example),
and the formation rate is less sensitive to the change in $\alpha_\chi$, i.e., it depends on $\alpha_\chi$ as power law instead of exponentially.
This explains why above the green line the $P=1$ (solid red) and $5\%$ (dashed red) curves deviate more from each other.

The force between any pair of DM particles is attractive and so for $m_{\phi}a_0\ll1$, the binding energy for a bound state with $N$ particles grows faster than linearly with $N$. That was what we found in the degenerate Fermi gas model where the binding energy grew as $N^{7/3}$. Thus it is likely that for $m_{\phi}<BE_0$ there are no thresholds that suppress the formation of non-relativistic bound states with more than two particles.

%

\section{Concluding remarks}

We showed that, for a range of parameters, one of the simplest low energy effective theories of asymmetric dark matter has a rich spectrum of bound states. In this paper, we explored some
of the features of the spectrum, and the implications for cosmology and dark matter direct detection. 
We find a region of parameter space where the dark matter in the universe is primarily in bound states.
Roughly speaking, this occurs when the binding energy of the two-particle ground state is greater than the mediator mass and the coupling of the mediator to the dark matter is large enough. 
We find that bound state formation and dissociation rates are suppressed because the operator mediating the transition is the unit operator in the dipole approximation and non relativistic limits. The matrix element is then the overlap of orthogonal wave functions which vanishes. Hence the transition matrix element for $\phi$ absorption and emission comes from small deviations from the dipole and non-relativistic approximations. Significant cosmological bound state production occurs only for rather large couplings, $\alpha_{\chi}\gtrsim0.1$. Later, after structures form there are other ways that bound states can form including in the core of neutron stars.

There are a number of issues that require further examination.
For example, the details of the spectrum of the bound states with more than two dark matter particles, i.e., nuggets, and the formation of these multi-particle bound states in the early universe. 
Without further investigation of these issues, it is even conceivable that for a range of parameters most of the asymmetric dark matter ends up as black holes.
In that case, the black holes must have a lifetime longer than the age of the universe.
It seems worthwhile to elucidate further the bound state properties and cosmology in this simple model for asymmetric dark matter.

Some of the work in this paper is also applicable to dark matter that is not asymmetric. For a range of parameters, scalar exchange could gives rise to stable bound states of dark matter particles and stable bound states of anti dark matter particles. One difference from the asymmetric case is that the values of $\alpha_\chi$ and $\alpha_I$ are constrained to give the correct dark matter relic density.

\section*{Acknowledgement}
We would like to thank Sean Carroll, Clifford Cheung, Francis-Yan Cyr-Racine, Michael Ramsey-Musolf, David Sanford, Philip Schuster and Natalia Toro for useful discussions and comments on a draft. YZ thanks Xiaoyong Chu for a comment on the calculation of the bound state formation rate.
This work is supported by the Gordon and Betty Moore Foundation through Grant
No.~776 to the Caltech Moore Center for Theoretical Cosmology and Physics, and by the DOE Grant DE-SC0011632,
and also by a DOE Early Career Award under Grant No. DE-SC0010255.

\section*{APPENDIX: A MORE NATURAL MODEL} \label{appx}

The smallness of $\mu_{\phi h}$ in the minimal we consider requires an awkward fine-tuning that is unlikely to have an environment origin.
More complicated models can avoid this feature. For example, suppose the dark sector possesses a global dark isospin symmetry $SU(2)$, under which
the DM $\chi$ is a doublet and the mediator $\phi$ is a triplet. The Lagrangian is
\begin{eqnarray}
\mathcal{L} &=&i\bar \chi \!\not\!\partial \chi - m_\chi \bar \chi \chi - g_\chi  \bar \chi \Phi \chi + \frac{1}{4} {\rm Tr} (\partial \Phi)^2 - \frac{1}{24}\lambda_{4\phi} {\rm Tr} \Phi^4 - \frac{1}{6}\lambda_{3\phi} {\rm Tr} \Phi^3 - \frac{1}{2} m_\phi^2 {\rm Tr} \Phi^2 \nonumber \\
& -& \frac{1}{2}\lambda_{\phi h} {\rm Tr} \Phi^2 (H^\dagger H-v^2/2) - V(H) \ ,
\end{eqnarray}
where $\Phi=\sigma^a \phi^a$. In order for the $\phi$ to decay, a dark doublet of left-handed fermions $\psi$ is introduced that couples to $\Phi$ via the interaction $\bar \psi^c \Phi \psi+{\rm h.c.}$. Note the dark isospin forbids a mass term for $\psi$. The coupling between $\Phi$ and $\psi$ should be large enough for $\Phi$ to decay before BBN.

In this model, the DM direct detection occurs at the one loop level (see Fig.~\ref{su2}). The cross section is given by Eq.~(\ref{directdetect}), with $\mu_{\phi h}/m_\phi^2$ replaced by
$\sim g_\chi \lambda_{\phi h}/(16\pi^2 m_\chi)$, which is adequately small even for $\lambda_{\phi h}$ of order unity.

\begin{figure}[h]
\includegraphics[width=0.35\columnwidth]{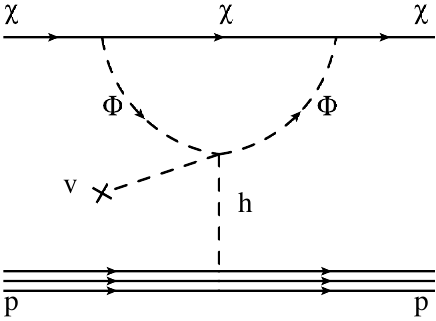}
\caption{Feynman diagram for direct detection in the $SU(2)$ model.}\label{su2}
\end{figure}

\end{document}